\documentclass[10pt,a4paper,onecolumn]{article}
\usepackage{marginnote}
\usepackage{graphicx}
\usepackage[rgb,svgnames]{xcolor}
\usepackage{authblk,etoolbox}
\usepackage{titlesec}
\usepackage{calc}
\usepackage{tikz}
\usepackage{orcidlink}
\let\JOSSorcidlink\orcidlink
\renewcommand{\orcidlink}[1]{\JOSSorcidlink{#1}\hspace{2pt}}
\usepackage{hyperref}
\hypersetup{colorlinks,breaklinks=true,
            urlcolor=[rgb]{0.0, 0.5, 1.0},
            linkcolor=Blue,
            citecolor=Blue}
\usepackage{caption}
\usepackage{tcolorbox}
\usepackage{amssymb,amsmath}
\usepackage{ifxetex,ifluatex}
\usepackage{seqsplit}
\usepackage{xstring}

\usepackage{float}
\let\origfigure\figure
\let\endorigfigure\endfigure
\renewenvironment{figure}[1][2] {
    \expandafter\origfigure\expandafter[H]
} {
    \endorigfigure
}

\usepackage{fixltx2e} 

\let\textttOrig=\texttt
\def\texttt#1{\expandafter\textttOrig{\seqsplit{#1}}}
\renewcommand{\seqinsert}{\ifmmode
  \allowbreak
  \else\penalty6000\hspace{0pt plus 0.02em}\fi}

\makeatletter
\let\href@Orig=\href
\def\href@Urllike#1#2{\href@Orig{#1}{\begingroup
    \def\Url@String{#2}\Url@FormatString
    \endgroup}}
\def\href@Notdoi#1#2{\def\tempa{#1}\def\tempb{#2}%
  \ifx\tempa\tempb\relax\href@Urllike{#1}{#2}\else
  \href@Orig{#1}{#2}\fi}
\def\href#1#2{%
  \IfBeginWith{#1}{https://doi.org}%
  {\href@Urllike{#1}{#2}}{\href@Notdoi{#1}{#2}}}
\makeatother

\NewDocumentCommand\citeproctext{}{}
\NewDocumentCommand\citeproc{mm}{%
  \hyperlink{cite.#1}{#2}}
\makeatletter
 \let\@cite@ofmt\@firstofone
 \def\@biblabel#1{}
 \def\@cite#1#2{{#1\if@tempswa , #2\fi}}
\makeatother
\newlength{\cslhangindent}
\newlength{\csllabelwidth}
\newenvironment{CSLReferences}[2] 
 {\begin{list}{}{%
  \setlength{\itemindent}{0pt}
  \setlength{\leftmargin}{0pt}
  \setlength{\parsep}{0pt}
  \ifodd #1
   \setlength{\leftmargin}{\cslhangindent}
   \setlength{\itemindent}{-1\cslhangindent}
  \fi
  \setlength{\itemsep}{#2\baselineskip}}}
 {\end{list}}
\usepackage{calc}

\usepackage[top=3.5cm, bottom=3cm, right=1.5cm, left=1.0cm,
            headheight=2.2cm, reversemp, includemp, marginparwidth=4.5cm]{geometry}

\titleformat{\section}
  {\normalfont\sffamily\Large\bfseries}
  {}{0pt}{}
\titleformat{\subsection}
  {\normalfont\sffamily\large\bfseries}
  {}{0pt}{}
\titleformat{\subsubsection}
  {\normalfont\sffamily\bfseries}
  {}{0pt}{}
\titleformat*{\paragraph}
  {\sffamily\normalsize}

\usepackage{fancyhdr}
\makeatletter
\let\ps@plain\ps@fancy
\fancyheadoffset[L]{4.5cm}
\fancyfootoffset[L]{4.5cm}

\definecolor{linky}{rgb}{0.0, 0.5, 1.0}

\newtcolorbox{repobox}
   {colback=red, colframe=red!75!black,
     boxrule=0.5pt, arc=2pt, left=6pt, right=6pt, top=3pt, bottom=3pt}

\newcommand{\ExternalLink}{%
   \tikz[x=1.2ex, y=1.2ex, baseline=-0.05ex]{%
       \begin{scope}[x=1ex, y=1ex]
           \clip (-0.1,-0.1)
               --++ (-0, 1.2)
               --++ (0.6, 0)
               --++ (0, -0.6)
               --++ (0.6, 0)
               --++ (0, -1);
           \path[draw,
               line width = 0.5,
               rounded corners=0.5]
               (0,0) rectangle (1,1);
       \end{scope}
       \path[draw, line width = 0.5] (0.5, 0.5)
           -- (1, 1);
       \path[draw, line width = 0.5] (0.6, 1)
           -- (1, 1) -- (1, 0.6);
       }
   }

\patchcmd{\@maketitle}{center}{flushleft}{}{}
\patchcmd{\@maketitle}{center}{flushleft}{}{}
\patchcmd{\@maketitle}{\LARGE}{\LARGE\sffamily}{}{}
\def\maketitle{{%
  
  \AB@maketitle}}
\makeatletter
\renewcommand\AB@affilsepx{ \protect\Affilfont}
\renewcommand\AB@affilnote[1]{{\bfseries #1}\hspace{3pt}}
\renewcommand{\affil}[2][]%
   {\newaffiltrue\let\AB@blk@and\AB@pand
      \if\relax#1\relax\def\AB@note{\AB@thenote}\else\def\AB@note{#1}%
        \setcounter{Maxaffil}{0}\fi
        \begingroup
        \let\href=\href@Orig
        \let\texttt=\textttOrig
        \let\protect\@unexpandable@protect
        \def\thanks{\protect\thanks}\def\footnote{\protect\footnote}%
        \@temptokena=\expandafter{\AB@authors}%
        {\def\\{\protect\\\protect\Affilfont}\xdef\AB@temp{#2}}%
         \xdef\AB@authors{\the\@temptokena\AB@las\AB@au@str
         \protect\\[\affilsep]\protect\Affilfont\AB@temp}%
         \gdef\AB@las{}\gdef\AB@au@str{}%
        {\def\\{, \ignorespaces}\xdef\AB@temp{#2}}%
        \@temptokena=\expandafter{\AB@affillist}%
        \xdef\AB@affillist{\the\@temptokena \AB@affilsep
          \AB@affilnote{\AB@note}\protect\Affilfont\AB@temp}%
      \endgroup
       \let\AB@affilsep\AB@affilsepx
}
\makeatother

\renewcommand\Affilfont{\sffamily\small\mdseries}
\ifnum 0\ifxetex 1\fi\ifluatex 1\fi=0 
  \usepackage[T1]{fontenc}
  \usepackage[utf8]{inputenc}

\else 
  \usepackage{fontspec}
  \defaultfontfeatures{Ligatures=TeX,Scale=MatchLowercase}
  \usepackage{fontsetup}
\fi
\IfFileExists{upquote.sty}{\usepackage{upquote}}{}
\IfFileExists{microtype.sty}{%
\usepackage{microtype}
\UseMicrotypeSet[protrusion]{basicmath} 
}{}

\usepackage{hyperref}
\hypersetup{unicode=true,
            pdftitle={ler: LVK (LIGO-Virgo-KAGRA collaboration) event (compact-binary mergers) rate calculator and simulator},
            pdfborder={0 0 0},
            breaklinks=true}
\let\addcontentslineOrig=\addcontentsline
\def\addcontentsline#1#2#3{\bgroup
  \let\texttt=\textttOrig\addcontentslineOrig{#1}{#2}{#3}\egroup}
\let\markbothOrig\markboth
\def\markboth#1#2{\bgroup
  \let\texttt=\textttOrig\markbothOrig{#1}{#2}\egroup}
\let\markrightOrig\markright
\def\markright#1{\bgroup
  \let\texttt=\textttOrig\markrightOrig{#1}\egroup}

\usepackage{graphicx,grffile}
\makeatletter
\def\maxwidth{\ifdim\Gin@nat@width>\linewidth\linewidth\else\Gin@nat@width\fi}
\def\maxheight{\ifdim\Gin@nat@height>\textheight\textheight\else\Gin@nat@height\fi}
\makeatother
\setkeys{Gin}{width=\maxwidth,height=\maxheight,keepaspectratio}
\IfFileExists{parskip.sty}{%
\usepackage{parskip}
}{
\setlength{\parindent}{0pt}
\setlength{\parskip}{6pt plus 2pt minus 1pt}
}
\ifx\paragraph\undefined\else
\let\oldparagraph\paragraph
\renewcommand{\paragraph}[1]{\oldparagraph{#1}\mbox{}}
\fi
\ifx\subparagraph\undefined\else
\let\oldsubparagraph\subparagraph
\renewcommand{\subparagraph}[1]{\oldsubparagraph{#1}\mbox{}}
\fi

\title{ler: LVK (LIGO-Virgo-KAGRA collaboration) event (compact-binary
mergers) rate calculator and simulator}

          \author[1]{Hemantakumar
Phurailatpam\,\protect\orcidlink{0000-0002-0471-3724}}
              \author[2,3]{Anupreeta
More\,\protect\orcidlink{0000-0001-7714-7076}}
              \author[4,5]{Harsh
Narola\,\protect\orcidlink{0000-0001-9161-7919}}
              \author[1,6]{Leo C. Y.
Ng\,\protect\orcidlink{0009-0000-3587-1832}}
              \author[4,5,7,8]{Justin
Janquart\,\protect\orcidlink{0000-0003-2888-7152}}
              \author[4,5]{Chris Van Den
Broeck\,\protect\orcidlink{0000-0001-6800-4006}}
              \author[1]{Otto Akseli
Hannuksela\,\protect\orcidlink{0000-0002-3887-7137}}
              \author[9]{Neha
Singh\,\protect\orcidlink{0000-0002-1135-3456}}
              \author[9]{David
Keitel\,\protect\orcidlink{0000-0002-2824-626X}}
      
      \affil[1]{Department of Physics, The Chinese University of Hong
Kong, Shatin, New Territories, Hong Kong}
      \affil[2]{The Inter-University Centre for Astronomy and
Astrophysics (IUCAA), Post Bag 4, Ganeshkhind, Pune 411007, India}
      \affil[3]{Kavli Institute for the Physics and Mathematics of the
Universe (IPMU), 5-1-5 Kashiwanoha, Kashiwa-shi, Chiba 277-8583, Japan}
      \affil[4]{Department of Physics, Utrecht University,
Princetonplein 1, 3584 CC Utrecht, The Netherlands}
      \affil[5]{Nikhef -- National Institute for Subatomic Physics,
Science Park, 1098 XG Amsterdam, The Netherlands}
      \affil[6]{IGC, Institute for Gravitation and the Cosmos,
Pennsylvania State University, University Park, PA 16802, USA}
      \affil[7]{Center for Cosmology, Particle Physics and Phenomenology
- CP3, Universit\textquotesingle e Catholique de Louvain,
Louvain-La-Neuve, B-1348, Belgium}
      \affil[8]{Royal Observatory of Belgium, Avenue Circulaire, 3, 1180
Uccle, Belgium}
      \affil[9]{IAC3, Universitat de les Illes Balears, Crta.
Valldemossa km 7.5, E-07122 Palma, Spain}
  \date{\vspace{-5.316ex}}

\begin{document}
\maketitle

\marginpar{

  \vspace{0.8255em}

  \begin{flushleft}
  \sffamily\small

  {\bfseries DOI:} \href{https://doi.org/10.21105/joss.07420}{\color{linky}{10.21105/joss.07420}}

  \vspace{2mm}

  {\bfseries Software}
  \begin{itemize}
    \setlength\itemsep{0em}
    \item \href{https://github.com/openjournals/joss-reviews/issues/7420}{\color{linky}{Review}} \ExternalLink
    \item \href{https://github.com/hemantaph/ler/}{\color{linky}{Repository}} \ExternalLink
    \item \href{https://doi.org/10.5281/zenodo.8087460}{\color{linky}{Archive}} \ExternalLink
  \end{itemize}

  \vspace{2mm}

  \par\noindent\hrulefill\par

  \vspace{2mm}

  {\bfseries Editor:} \href{https://warrickball.gitlab.io/}{Warrick Ball} \ExternalLink
  \,\orcidlink{0000-0002-4773-1017} \\
  \vspace{1mm}
    {\bfseries Reviewers:}
  \begin{itemize}
  \setlength\itemsep{0em}
    \item \href{https://github.com/sibirrer}{@sibirrer}
    \item \href{https://github.com/Jammy2211}{@Jammy2211}
    \end{itemize}
    \vspace{2mm}

  {\bfseries Submitted:} 16 August 2024\\
  {\bfseries Published:} 16 August 2026

  \vspace{2mm}
  {\bfseries License}\\
  Authors of papers retain copyright and release the work under a Creative Commons Attribution 4.0 International License (\href{https://creativecommons.org/licenses/by/4.0/}{\color{linky}{CC BY 4.0}}).

  \end{flushleft}
}

\vspace{1.5em}

\section{Summary}\label{summary}

Gravitational waves (GWs) are ripples in spacetime produced by
accelerating massive objects. Since 2015, the LIGO-Virgo-KAGRA (LVK)
collaboration (\citeproc{ref-VIRGO:2015}{Acernese et al., 2014};
\citeproc{ref-KAGRA:2021}{Akutsu et al., 2020};
\citeproc{ref-LIGO:2015}{The LIGO Scientific Collaboration et al.,
2015}) has routinely detected GW transients from compact-binary mergers,
including binary black holes (BBHs), binary neutron stars (BNSs), and
neutron star-black hole pairs (NSBHs), during observing runs
(\citeproc{ref-GWTC3:2023}{R. Abbott et al., 2023};
\citeproc{ref-GWTC4:2026}{The LIGO Scientific Collaboration, the Virgo
Collaboration, and the KAGRA Collaboration, 2026}; R. Abbott et al.
\citeproc{ref-GWTC2:2021}{2021}, \emph{PRX};
\citeproc{ref-GWTC1:2019}{B. P. Abbott et al., 2019}). Most of these
signals reach detectors unobstructed, forming a baseline unlensed
population, but a small fraction can encounter massive intervening
objects such as galaxies or galaxy clusters. These objects act as
lenses, splitting the signals into multiple images that are magnified,
time-shifted, and phase-shifted through strong gravitational lensing (R.
Abbott et al. \citeproc{ref-Abbott:2021}{2021}, \emph{ApJ};
\citeproc{ref-Wierda:2021}{Wierda et al., 2021}).

Accurate modeling of unlensed and lensed gravitational-wave populations
is essential for astrophysical interpretation, inferring merger rate
histories, and validating future lensing events (R. Abbott et al.
\citeproc{ref-Abbott:2021}{2021}, \emph{ApJ};
\citeproc{ref-Janquart:2023}{Janquart et al., 2023};
\citeproc{ref-ligolensing:2024}{The LIGO Scientific Collaboration, the
Virgo Collaboration, and the KAGRA Collaboration, 2024};
\citeproc{ref-Wierda:2021}{Wierda et al., 2021}). The \texttt{ler}
software is a Python framework that provides a unified pipeline to model
these populations statistically and calculate their detectable rates.
Producing precise results requires computationally demanding tasks such
as large-scale sampling of source and lens attributes, solving lens
equations, and simulating detector responses. \texttt{ler} reduces these
costs through optimization and parallelization, achieving up to a
thousandfold speedup relative to an un-optimized baseline implementation
(further details provided in the Software design section). Source code,
validation examples, and documentation are available through the project
repository, \href{https://github.com/hemantaph/lertest}{supporting
examples}, and \href{https://ler.hemantaph.com}{documentation site}.

\section{Statement of need}\label{statement-of-need}

\texttt{ler} is a statistics-based Python tool designed for the joint
simulation and computation of detectable rates for both unlensed and
lensed gravitational-wave events. It is intended for LVK analyses and
for astrophysics researchers working on compact-binary populations and
strong lensing. Its core functionality integrates the sampling of
compact-binary source properties and lens-galaxy characteristics,
solving lens equations to derive image properties, and computing
detectable event rates. This functionality leverages NumPy
(\citeproc{ref-numpy:2020}{Harris et al., 2020}) for array operations
and linear algebra alongside SciPy (\citeproc{ref-scipy:2020}{Virtanen
et al., 2020}) interpolation methods and Python multiprocessing.
Efficiency is improved using Numba (\citeproc{ref-numba:2015}{Lam et
al., 2015}) for just-in-time compilation and multithreading, and by
using inverse-transform sampling and importance sampling where rejection
sampling is inefficient. \texttt{ler} can use \texttt{lenstronomy}
(\citeproc{ref-Birrer:2021}{Birrer et al., 2021}) for lensing
calculations and \texttt{gwsnr}
(\citeproc{ref-Phurailatpam:2025}{Phurailatpam \& Hannuksela, 2025}) for
efficient detection-probability calculations \(P_{\mathrm{det}}\).

\texttt{ler} produces parameter distributions for intrinsic populations
and for populations selected by detection criteria, including
strong-lensing selection and image-level detectability. This supports
validation studies for lensing candidates
(\citeproc{ref-Janquart:2023}{Janquart et al., 2023}) and forecasting
for present and future detector networks (\citeproc{ref-Ng:2025}{Ng et
al., 2025}). The design can be adapted to multi-messenger applications
by changing the detection criteria, including studies of strongly lensed
gamma-ray bursts (\citeproc{ref-More:2025}{More \& Phurailatpam, 2025}).
\texttt{ler} also supports integration with existing inference tools by
providing \(P_{\mathrm{det}}\) estimates that can be used to construct
selection functions (see
\href{https://github.com/hemantaph/lertest}{supporting examples}) for
hierarchical population inference frameworks
(\citeproc{ref-Thrane:2019}{Thrane \& Talbot, 2019}) with Bilby software
(\citeproc{ref-Ashton:2019}{Ashton et al., 2019}).

\section{State of the field}\label{state-of-the-field}

Joint population studies of unlensed and strongly lensed gravitational
waves require end-to-end simulations that combine source and lens-galaxy
models with detector selection effects. Common inference packages such
as Bilby (\citeproc{ref-Ashton:2019}{Ashton et al., 2019}) and cosmology
and rate tools such as
\href{https://git.ligo.org/lscsoft/gwcosmo}{\texttt{gwcosmo}}
(\citeproc{ref-Gray:2020gwcosmo}{Gray et al., 2020},
\citeproc{ref-Gray:2022gwcosmo}{2022},
\citeproc{ref-Gray:2023gwcosmo}{2023}) provide general-purpose
likelihood evaluation, sampling, and population inference utilities.
However, they do not provide a single, configuration-driven pipeline to
generate unlensed and strongly lensed populations jointly at scale, nor
to propagate lensing image properties into detectability. Lens-modelling
tools such as \texttt{lenstronomy} (\citeproc{ref-Birrer:2021}{Birrer et
al., 2021}) implement fast analytical routines for lensing calculations
but are not designed as high-throughput population-level simulators.

Previous strong-lensing population studies often relied on pipelines
tightly coupled to specific lens models or detection criteria
(\citeproc{ref-Haris:2018}{Haris et al., 2018};
\citeproc{ref-Wierda:2021}{Wierda et al., 2021}). Many studies report
rate estimates based on limited sample sizes without showing convergence
diagnostics or uncertainty estimates. They also commonly neglect
uncertainty in population-model hyperparameters, which should propagate
into the detectable population and rate predictions. \texttt{ler}
addresses these gaps by providing an integrated, high-throughput
workflow linking population draws, lensing cross-sections, image
properties, and detector selection in one reproducible pipeline.

\section{\texorpdfstring{Software design
\label{sec:software-design}}{Software design }}\label{software-design}

\texttt{ler} is designed as an end-to-end population simulator with a
clear separation between the scientific steps and the computational
machinery. The pipeline is organized around two rate calculations.
Unlensed rates are computed by drawing compact-binary populations from
astrophysical priors, evaluating detectability, and integrating over the
detected events. Lensed rates extend this flow by sampling lens-galaxy
and source-redshift parameters under a strong-lensing condition, solving
the lens equation to generate image properties, applying detectability
criteria at the image level, and integrating over the detectable lensed
events. This separation keeps the rate logic explicit while allowing the
source, lens, and detection models to be replaced without rewriting the
full pipeline. The mathematical foundations for these implementations
are detailed in the Mathematics section.

A core design choice is modularity, with stable interfaces between
modules that handle source populations, lens populations, image-property
calculations, and rate estimation. Users can select built-in models such
as BBH, BNS, or NSBH populations and lens models such as singular
isothermal sphere (SIS), singular isothermal ellipsoid (SIE), or
elliptical power-law with external shear (EPL+Shear), or provide their
own distributions and detection criteria. Detector selection is handled
through a backend interface so that signal-to-noise and detection
probability calculations can be delegated to \texttt{gwsnr} or replaced
by alternative implementations. This architecture supports extension and
testing while keeping the scientific assumptions visible in
configuration rather than embedded in ad hoc scripts.

Performance is obtained by compiling the repeatedly evaluated parts of
the pipeline with Numba \texttt{njit} and parallelizing the main Monte
Carlo loops. \texttt{ler} uses \texttt{lenstronomy}
(\citeproc{ref-Birrer:2021}{Birrer et al., 2021}) as an optional
backend, but it provides an in-house analytical EPL+Shear solver and
caustic and cross-section calculator that are compiled with Numba and
run in parallel, following the same mathematical formulation as
\texttt{lenstronomy} and Tessore \& Benton Metcalf
(\citeproc{ref-Tessore:2015}{2015}). This leads to two practical
trade-offs. First, the rate estimates are Monte Carlo approximations, so
accuracy is controlled by sample size rather than closed-form
expressions, and \texttt{ler} reports convergence statistics from
repeated batches. Second, just-in-time compilation adds overhead on
first use, which is about 2 s for the first unlensed batch and about 25
s for the first lensed batch that includes image-property calculations.
After compilation, the compiled kernels reduce per-sample cost and make
large simulations feasible. In benchmarks that disable just-in-time
compilation and parallelism, these choices provide speed gains of about
\(180\times\) for unlensed populations and \(5189\times\) for lensed
populations. On a standard six-core processor (Apple M2 Pro, 16GB RAM),
a batch of \(10^5\) samples takes about 300 ms for unlensed calculations
and about 30 s for lensed calculations, and repeated batches converge to
a stable rate estimate.

\section{Mathematics}\label{mathematics}

The mathematical workflow in \texttt{ler} can be summarised as Monte
Carlo estimates of detectable event rates for unlensed and strongly
lensed populations, obtained by averaging detection probabilities over
simulated events drawn from the corresponding population distributions.
For unlensed events, the detector-frame rate of detectable events is

\[
\frac{\Delta N^{\mathrm{obs}}_{\mathrm{U}}}{\Delta t}
=
\mathcal{N}_{\mathrm{U}}
\bigg\langle
P(\mathrm{obs}\mid\vec{\theta}_{\mathrm{U}})
\bigg\rangle_{\vec{\theta}_{\mathrm{U}}\sim P(\vec{\theta}_{\mathrm{U}})}
\, ,
\]

where \(\mathcal{N}_{\mathrm{U}}\) is the total intrinsic merger rate in
the detector frame, \(\vec{\theta}_{\mathrm{U}}\) denotes the
15-dimensional set of gravitational-wave source parameters drawn from
\(P(\vec{\theta}_{\mathrm{U}})\), and
\(P(\mathrm{obs}\mid\vec{\theta}_{\mathrm{U}})\) is the detection
probability for a source with parameters \(\vec{\theta}_{\mathrm{U}}\).

For strongly lensed events, the detector-frame rate of detectable lensed
events is

\[
\frac{\Delta N^{\mathrm{obs}}_{\mathrm{L}}}{\Delta t}
=
\mathcal{N}_{\mathrm{L}}
\bigg\langle
P(\mathrm{obs}\mid
\vec{\theta}_{\mathrm{U}},
\vec{\theta}_{\mathrm{L}},
\vec{\beta},
\mathrm{SL})
\bigg\rangle_{
\substack{
\vec{\theta}_{\mathrm{U}},\vec{\theta}_{\mathrm{L}}
\sim
P(\vec{\theta}_{\mathrm{U}},\vec{\theta}_{\mathrm{L}}\mid\mathrm{SL})
\\
\vec{\beta}
\sim
P(\vec{\beta}\mid
\vec{\theta}_{\mathrm{U}},\vec{\theta}_{\mathrm{L}},\mathrm{SL})
}}
\, ,
\]

where \(\mathcal{N}_{\mathrm{L}}\) is the total strongly lensed event
rate in the detector frame before detector selection,
\(\vec{\theta}_{\mathrm{L}}\) denotes the lens-related parameters, and
\(\vec{\beta}\) is the source position in the source plane. The
distributions
\(P(\vec{\theta}_{\mathrm{U}},\vec{\theta}_{\mathrm{L}}\mid\mathrm{SL})\)
and
\(P(\vec{\beta}\mid\vec{\theta}_{\mathrm{U}},\vec{\theta}_{\mathrm{L}},\mathrm{SL})\)
define the hierarchical sampling conditioned on strong lensing, denoted
by \(\mathrm{SL}\). For each source-lens configuration and source
position,
\(P(\mathrm{obs}\mid\vec{\theta}_{\mathrm{U}},\vec{\theta}_{\mathrm{L}},\vec{\beta},\mathrm{SL})\)
gives the probability that the resulting lensed event satisfies the
adopted detection criterion. Its evaluation requires solving the lens
equation to obtain the image properties, applying the corresponding
lensing transformations to the GW signal, evaluating the detection
criterion for each image, and requiring at least two images to be
detectable for the lensed event to satisfy the event-level detection
criterion.

\section{Research impact statement}\label{research-impact-statement}

\texttt{ler} has established a measurable impact within the
gravitational-wave community through its integration into research
workflows and its growing user base. The package supports a diverse
community ranging from students to faculty members at multiple
institutions, with development driven by researchers across various
organizations and
\href{https://dcc.ligo.org/LIGO-P2400274}{peer-reviewed within the LVK
Scientific Collaboration}. Active community engagement is maintained
through bug reports and feature requests on the GitHub issue tracker and
through collaboration communication channels such as LIGO Mattermost.

The software has facilitated several scientific publications related to
gravitational-wave lensing statistics. It has been used to analyze the
distribution of time delays and magnification ratios for the validation
of lensing candidates (\citeproc{ref-Janquart:2023}{Janquart et al.,
2023}) and to forecast the detection of sub-threshold lensed events with
future observatories (\citeproc{ref-Ng:2025}{Ng et al., 2025}).
Additional research applications include the generation of lensed-event
statistics and the calculation of posterior odds
(\citeproc{ref-Hannuksela:2026}{Hannuksela et al., 2026}), as well as
forecasting electromagnetic counterparts for strongly lensed gamma-ray
bursts (\citeproc{ref-More:2025}{More \& Phurailatpam, 2025}). These
implementations demonstrate that the tool provides a robust framework
for high-level population studies and astrophysical forecasting.

\section{AI usage disclosure}\label{ai-usage-disclosure}

No generative AI tools were used to develop the software, write this
manuscript, or prepare the accompanying materials.

\section{Acknowledgements}\label{acknowledgements}

The authors thank their academic advisors for guidance and support.
Hemantakumar Phurailatpam acknowledges the Department of Physics at The
Chinese University of Hong Kong for the Postgraduate Studentship that
facilitated this research. Hemantakumar Phurailatpam and Otto A.
Hannuksela acknowledge support from the Research Grants Council of Hong
Kong (Project Nos. CUHK 14304622 and 14307923), the start-up grant from
The Chinese University of Hong Kong, and the Direct Grant for Research
from the Research Committee of The Chinese University of Hong Kong. The
authors also acknowledge support from the Netherlands Organisation for
Scientific Research (NWO). N. Singh is supported through the Conselleria
d'Educació i Universitats del Govern de les Illes Balears via a Vicenç
Mut postdoctoral grant (POSTDOC2024\_55) with funds from the European
Social Fund+ in the framework of the Balearic Islands ESF+ Program
2021-2027. N. Singh and D. Keitel are supported by the Universitat de
les Illes Balears (UIB) with funds from the Programa de Foment de la
Recerca i la Innovació de la UIB 2024-2026 (supported by the yearly plan
of the Tourist Stay Tax ITS2023-086); the Spanish Agencia Estatal de
Investigación grants PID2022-138626NB-I00, RED2024-153978-E,
RED2024-153735-E, funded by MICIU/AEI/10.13039/501100011033 and the
ERDF/EU; and the Comunitat Autònoma de les Illes Balears through the
Conselleria d'Educació i Universitats with funds from the European Union
- European Regional Development Fund (ERDF) (SINCO2022/18146 -
Plataforma HiTech-IAC3-BIO). The authors acknowledge the computational
resources provided by the LIGO Laboratory, supported by National Science
Foundation Grants No.~PHY-0757058 and No.~PHY-0823459.

\section*{References}\label{references}
\addcontentsline{toc}{section}{References}

\phantomsection\label{refs}

\end{document}